\documentclass[12pt,a4paper]{article}
\usepackage{graphicx}
\usepackage[T1]{fontenc}
\usepackage[utf8]{inputenc}
\usepackage{textcomp}
\usepackage[sc,osf]{mathpazo}
\usepackage{a4wide}  
\usepackage{latexsym,amsthm,amsfonts,amsmath,mathrsfs,amssymb}
\usepackage{dsfont}
\usepackage{accents}
\usepackage[nosort]{cite}
\usepackage{booktabs} 
\usepackage[unicode,implicit]{hyperref}
\hypersetup{%
  pdftitle    = {The Komar charge in presence of the Holst term
    and the gravitational Witten effect}
  pdfkeywords = {gravity, Palatini, Holst, Einstein-Hilbert, conserved
    quantities, Komar charge, Taub-NUT solution, NUT charge,
    electric-magnetic duality, Witten effect},
  pdfauthor   = {Jos\'e Luis V. Cerdeira and Tom\'as Ort\'{\i}n},
  plainpages  = true,
  colorlinks  = true,
  citecolor   = blue,
  urlcolor    = red,
  linkcolor   = black
}
\newcommand{\hepth}[1]{{\tt
\href{http://www.arXiv.org/abs/hep-th/#1}{hep-th/#1}}}
\newcommand{\grqc}[1]{{\tt
\href{http://www.arXiv.org/abs/gr-qc/#1}{gr-qc/#1}}}

\newcommand{\arxiv}[1]{{\tt arXiv:\href{http://www.arXiv.org/abs/#1}{#1}}}
\newcommand{\doi}[1]{{\tt DOI:\href{http://dx.doi.org/#1}{#1}}}

\allowdisplaybreaks

\begin{document}

\begin{flushright}
\small
IFT-UAM/CSIC-25-059\\
June 5\textsuperscript{th}, 2025\\
\normalsize
\end{flushright}

\vspace{.2cm}

\begin{center}

  {\Large {\bf The Komar charge in presence of the Holst term\\[.5cm]
    and the gravitational Witten effect}}
 
\vspace{1cm}

\renewcommand{\thefootnote}{\alph{footnote}}

{\sl Jos\'e Luis V.~Cerdeira}$^{1,}$\footnote{Email: {\tt  jose.verez-fraguela[at]estudiante.uam.es}}
{\sl and Tom\'{a}s Ort\'{\i}n}$^{2,}$\footnote{Email: {\tt Tomas.Ortin[at]csic.es}}

\setcounter{footnote}{0}
\renewcommand{\thefootnote}{\arabic{footnote}}
\vspace{1cm}

${}^{1}${\it\small Instituto de F\'{\i}sica Corpuscular (IFIC), University of
  Valencia-CSIC,\\
Parc Cient\'{\i}fic UV, C/ Catedr\'atico Jos\'e Beltr\'an 2, E-46980 Paterna, Spain}

\vspace{0.2cm}

${}^{2}${\it\small Instituto de F\'{\i}sica Te\'orica UAM/CSIC\\
C/ Nicol\'as Cabrera, 13--15,  C.U.~Cantoblanco, E-28049 Madrid, Spain}

\vspace{1cm}

{\bf Abstract}
\end{center}
\begin{quotation}
  {\small In the first-order formalism, the Einstein--Hilbert action can be
    modified by the addition of a Holst term multiplied by the Barbero
    parameter $\alpha$. This modification breaks parity although it does not
    affect the equations of motion. We show that the standard Komar charge is
    also modified by the addition of a topological term multiplied by the
    Barbero parameter $\alpha$. For the Killing vector that generates time
    translations, the value of the Komar integral at infinity is modified by
    the addition of a term proportional to the NUT charge $N$ and the
    parity-breaking Barbero parameter. Thus, as in the standard Witten
    effect, a non-vanishing NUT charge $N$ induces a non-vanishing mass
    $-\alpha N$.
  }
\end{quotation}

\newpage
\pagestyle{plain}
\noindent
\textbf{1. Introduction.} The second-order Einstein--Hilbert (EH) action is
the simplest non-trivial diff-invariant action one can write for the
gravitational field. It leads to equations of second order in derivatives of
the gravitational field (the Einstein equations) and it is natural to ask if
it can be modified adding other diff-invariant terms while preserving the
second-order degree of the equations of motion.

As Einstein himself soon discovered, the simplest modification that satisfies
this condition is the addition of the cosmological-constant term. As a matter
of fact, in $d=4$ dimensions, the only other diff-invariant terms that one can
add are either topological (the Gauss--Bonnet and the Euler 4-forms) or
boundary terms that do not modify the equations of motion
\cite{Lovelock:1971yv}.

The first-order (Palatini \cite{Palatini:1919ffw}) formalism turns out to
offer more possibilities because one can use terms that would vanish for the
Levi--Civita connection but do not vanish off-shell, for a generic
connection. In $d=4$ dimensions, these are the Nieh--Yan
\cite{Nieh:1981ww,Nieh:2008btw} and Holst (H) \cite{Holst:1995pc} terms. The
first is quadratic in torsion and vanishes for the torsionless Levi--Civita
connection. The second is proportional to a term that becomes the Bianchi
identity of the Riemann curvature (\textit{i.e.}~when the connection is the
Levi--Civita connection). They differ by a total derivative.

Although, out of all these possible additions, only the cosmological term
modifies the Einstein equations, all of them give non-trivial contributions to
the Noether--Wald charge.\footnote{See Ref.~\cite{Corichi:2016zac} for a
  comprehensive and pedagogical review. On the other hand, using the ``gravity
  as a gauge theory of the Poincar\'e group'' approach one can show that the
  list of terms one can add to the first-order action in $d=4$ dimensions that
  we have given above is exhaustive.} In this letter we will only consider the
H terms, not just for the sake of simplicity, but because, as we are going to
show, the H term is the analogue of the parity-breaking, topological,
$\theta$-term, $\theta F\wedge F$, that one can add to the Maxwell Lagrangian
$F\wedge \star F$. As we are going to show, this similarity between the H term
in gravity and the $\theta$ term in electromagnetism implies that their
presence also has similar effects.\footnote{Interestingly, the Host term can
  be supersymmetrized \cite{Kaul:2007gz,Szczachor:2012hh,Tomova:2024cgz}. In
  the last of these references it was also shown that, in the supergravity
  context, the NUT charge is associated to the supersymmetrization of the
  Holst term.}

\vspace{.3cm}
\noindent
\textbf{2. The Einstein--Hilbert and Holst terms.} In the first-order
formulation, the Vielbein 1-form $e^{a}\equiv e^{a}{}_{\mu}dx^{\mu}$ and the
Lorentz connection
$\omega^{ab} \equiv \omega_{\mu}{}^{ab}dx^{\mu}=-\omega^{ba}$ are independent
variables. The Lorentz curvature 2-form is in our conventions\footnote{Our
  conventions are those of Ref.~\cite{Ortin:2015hya}. Notice that the Lorentz
  connection with the Minkowski metric in tangent space $\eta_{ab}$. Thus, if
  its torsion
  $T^{a}\equiv -\mathcal{D}e^{a} =-\left(de^{a}-\omega^{a}{}_{b}\wedge
    e^{b}\right)$ vanishes, it is the Levi--Civita connection and the Lorentz
  curvature 2-form is equal to the Riemann tensor
  $R_{\mu\nu}{}^{ab} =
  R_{\mu\nu}{}^{\rho\sigma}e^{a}{}_{\rho}e^{b}{}_{\sigma}$.}
\begin{equation}
  R^{ab}
  =
  d\omega^{ab} -\omega^{a}{}_{c}\wedge \omega^{cb}.
\end{equation}

In differential-form language, the 4-dimensional first-order EH action can be
written in the form
\begin{equation}
\label{eq:EH}
S_{\rm EH}[e,\omega]
=
-\frac{1}{16\pi G_{\rm N}^{(4)}}\int_{\mathcal{M}}
e^{a}\wedge e^{b}\wedge  \tilde{R}_{ab}(\omega),
\end{equation}
where $G_{\rm N}^{(4)}$ is the 4-dimensional Newton constant and where we have
defined
\begin{equation}
  \tilde{R}_{ab}
  \equiv
  \tfrac{1}{2}\varepsilon_{abcd}R^{cd}.   
\end{equation}

Under an arbitrary variation of the Vielbein and connection
\begin{equation}
  \delta S
  =
  \int
  \left\{
    \mathbf{E}_{a}\wedge \delta e^{a}
    +\mathbf{E}_{ab}\wedge \delta\omega^{ab}
    +d\mathbf{\Theta}
  \right\},
\end{equation}
where
\begin{subequations}
  \begin{align}
  \mathbf{E}_{a}
    & =
      -2 e^{b}\wedge  \tilde{R}_{ba},
    \\
    \mathbf{E}_{ab}
    & =
      -\varepsilon_{cdab}T^{c}\wedge e^{d},
    \\
    \mathbf{\Theta}
    & =
      -\tfrac{1}{2}\varepsilon_{cdab}e^{c}\wedge e^{d}\wedge \delta\omega^{ab}.
  \end{align}
\end{subequations}

The equations of the Vielbein $\mathbf{E}_{a}=0$ imply that the Einstein
tensor vanishes. These are not yet the standard Einstein equations. The
equations of the connection $\mathbf{E}_{ab}=0$ imply that the torsion
vanishes.\footnote{In components, they read
  \begin{equation}
    \varepsilon_{cdab}T_{[ef}{}^{c}\eta_{g]}{}^{d}
    =
    0.
  \end{equation}
  Contracting with $\varepsilon^{efgh}$ they reduce to
  \begin{equation}
    T_{[cd}{}^{a}\eta_{a]}{}^{h}
    =
    0.
  \end{equation}
  Setting $d=h$ we get $T_{cd}{}^{d}=0$ and substituting back into the above
  equation this result we get $T=0$.  } Therefore, the connection is the
Levi--Civita connection and, upon the use of this solution in the equations of
the Vielbein, we recover the Einstein equations.

The H action \cite{Holst:1995pc} was originally introduced to provide a
Lagrangian origin for the Hamiltonian variables introduced by Barbero in
Ref.~\cite{BarberoG:1994eia} to solve some of the problems of Ashtekar's
variables \cite{Ashtekar:1986yd,Ashtekar:1987gu}. It can be obtained by
replacing in the EH $R_{ab}$ by $\tilde{R}_{ab}$
($\tilde{\tilde{R}}_{ab}=-R_{ab}$)
\begin{equation}
\label{eq:H}
S_{\rm EH}[e,\omega]
=
\frac{1}{16\pi G_{\rm N}^{(4)}}\int_{\mathcal{M}}
e^{a}\wedge e^{b}\wedge  R_{ab}(\omega),
\end{equation}
and it leads to
\begin{subequations}
  \begin{align}
  \mathbf{E}_{a}
    & =
      2 e^{b}\wedge R_{ba},
    \\
    \mathbf{E}^{ab}
    & =
      2T^{[a}\wedge e^{b]},
    \\
    \mathbf{\Theta}
    & =
      e^{a}\wedge e^{b}\wedge \delta\omega_{ab}.
  \end{align}
\end{subequations}

The equation of the connection has the same solution as in the EH action but,
substituting into the Vielbein equations, these (and the whole H action)
vanish identically by virtue of the Bianchi identity\footnote{This identity
  follows trivially from the Ricci identity
  \begin{equation}
    [\mathcal{D},\mathcal{D}]e^{a}
    =
    -R^{a}{}_{b}\wedge e^{b},  
  \end{equation}
  and from the absence of torsion.
}
\begin{equation}
  \label{eq:Bianchiidentity}
e^{b}\wedge R_{ba} =0.  
\end{equation} 

As noticed in Ref.~\cite{Kol:2022bsd}, the replacement
$R^{ab}\to\tilde{R}^{ab}$ interchanges equations of motion and Bianchi
identities and can be seen as gravitational electric-magnetic duality
transformation. It also changes the parity of the action: the EH action is
parity-even while the Holst action is parity-odd. If the EH term is seen as
the kinetic term of the gravitational field, it is natural to regard the H
term as the analog of the $\theta$-term in the Yang--Mills theory. They can
only be defined in 4 dimensions.

In the first-order formalism we can combine these two actions, breaking
parity, introducing the so-called \textit{Barbero parameter}
$\alpha$\footnote{Notice that
  \begin{equation}
    \eta_{ab\, cd}
    =
    \tfrac{1}{2}(\eta_{ac}\eta_{bd}-\eta_{ad}\eta_{bc}),
    \,\,\,\,\,
    \Rightarrow
    \,\,\,\,\,
    \eta_{ab\, cd} e^{a}\wedge e^{b}\wedge  R^{cd}
    =
    e^{a}\wedge e^{b}\wedge  R_{ab}.
  \end{equation}
} 
\begin{equation}
  \label{eq:actiongaugePoincared4}
  \begin{aligned}
  S[e,\omega]
  & =
  \frac{1}{16\pi G_{N}^{(4)}}
  \int
  \left\{
    -\tfrac{1}{2}\varepsilon_{abcd}e^{a}\wedge e^{b}\wedge R^{cd}
    +\alpha\eta_{ab\, cd} e^{a}\wedge e^{b}\wedge R^{cd}
\right\}.
  \end{aligned}
\end{equation}

As we are going to see, this action leads, again, to the Einstein
equations. However, the H term will modify the definition of some of the
conserved charges and our goal is to find the Komar charge for this combined
action. To this end, we are going to use the standard techniques, summarized,
for instance, in Ref.~\cite{kn:JLVCTO}

\vspace{.3cm}
\noindent
\textbf{3. The Komar charge.} Under a generic variation of the fields, the
action Eq.~(\ref{eq:actiongaugePoincared4}) transforms as
\begin{equation}
    \label{eq:genericvariationactiongaugePoincared4}
  \delta S
  =
  \int \left\{ \mathbf{E}_{a}\wedge \delta e^{a} +\mathbf{E}_{ab}\wedge \delta
  \omega^{ab} +d\mathbf{\Theta}(e, \omega,\delta \omega)\right\},
\end{equation}
where, defining
\begin{equation}
  X_{abcd}
  \equiv
  -\tfrac{1}{2}\varepsilon_{abcd}+\alpha \eta_{ab\, cd},
\end{equation}
and, ignoring the overall normalization factor
$\left(16\pi G_{N}^{(4)}\right)^{-1}$, the equations of motion of the Vielbein
and spin connection are
\begin{subequations}
  \begin{align}
    \mathbf{E}_{a}
    & =
      -2 X_{abcd}e^{b}\wedge R^{cd},
    \\
    \mathbf{E}_{ab}
    & =
      2 X_{abcd}T^{c}\wedge e^{d},
  \end{align}
\end{subequations}
while the presymplectic potential is given by
\begin{equation}
  \mathbf{\Theta}(e, \omega,\delta \omega)
  =
  X_{abcd}e^{a}\wedge e^{b}\wedge \delta \omega^{cd}.
\end{equation}

The equation $\mathbf{E}_{ab}=0$ is solved by
$T^{[a}\wedge e^{b]}=0$\footnote{Contracting it with $\varepsilon^{abef}$ on
  obtains a similar equation with different coefficients multiplying
  $\eta_{ab\, cd}$ and $\varepsilon_{abcd}$. The new equation can be combined
  with the old one to get that result.} which, as we have see, is solved, in
its turn, by $T^{a}=0$ so that the connection is, again, Levi--Civita
connection.  If we substitute this result into the Vielbein equations
$\mathbf{E}_{a}$, the term coming from the H action vanishes identically by
virtue of the Bianchi identity Eq.~(\ref{eq:Bianchiidentity}) and we end up
with the usual Einstein equations.

The H term in the symplectic prepotential does not vanish, though, and it
will generate a new term in the Komar charge.

The action Eq.~(\ref{eq:actiongaugePoincared4}) is invariant under general
coordinate transformations up to a total derivative:
\begin{equation}
\delta_{\xi}S = -\int \imath_{\xi}\mathbf{L}.  
\end{equation}
The left-hand side of this equation can be computed particularizing the
general variation of the action in
Eq.~(\ref{eq:genericvariationactiongaugePoincared4})
\begin{equation}
  \delta_{\xi} S
  =
  \int \left\{ \mathbf{E}_{a}\wedge \delta_{\xi} e^{a}
    +\mathbf{E}_{ab}\wedge \delta_{\xi}\omega^{ab}
    +d\mathbf{\Theta}(e, \omega,\delta_{\xi} \omega)\right\},
\end{equation}
and using in it the transformations of the Vielbein and connection
$\delta_{\xi}e^{a}$ and $\delta_{\xi}\omega^{ab}$. As explained, for instance,
in Ref.~\cite{Elgood:2020svt}, these transformations must take into account
the induced (or compensating) local Lorentz transformations and, taking into
account the possible torsion of the connection as in
Refs.~\cite{Elgood:2020nls,Bandos:2023zbs,Bandos:2024rit},\footnote{We review
  the derivation of these transformations for the current setting in the
  Appendix. We stress that the connection in the covariant derivative
  $\mathcal{D}$ has torsion.} can be written in the form
Eq.~(\ref{eq:deltaxiea}) and (\ref{eq:deltaxioab}), which we rewrite here for
convenience
\begin{subequations}
  \begin{align}
    \delta_{\xi}e^{a}
    & =
      -\left(\mathcal{D}\xi^{a}  -\imath_{\xi}T^{a}
      +P_{\xi}{}^{a}{}_{b}e^{b}\right),
    \\
      \delta_{\xi}\omega^{ab}
    & =
      -\left(\imath_{\xi}R^{ab}+\mathcal{D}P_{\xi}{}^{ab}\right),
  \end{align}
\end{subequations}
where $P_{\xi}{}^{ab}$ is a local Lorentz parameter given in
Eq.~(\ref{eq:momentummap})  that we also write here for convenience:
\begin{equation}
    P_{\xi}{}^{ab}
  =
  \mathcal{D}^{[a}\xi^{b]} -\xi^{c}T_{c}{}^{[ab]}.
\end{equation}

As shown in the Appendix, these definitions guarantee that, when $\xi$ is a
Killing vector $k$ (\textit{i.e.}~it leaves the metric invariant) that also
leaves the torsion field invariant, the above transformations vanish
identically.

We get
\begin{equation}
  \begin{aligned}
  \delta_{\xi} S
    & =
      \int \left\{ -\xi^{a}\left[\mathcal{D}\mathbf{E}_{a}
      -\mathbf{E}_{c}\wedge \imath_{a}T^{c}
      +\mathbf{E}_{bc}\wedge \imath_{a}R^{bc}\right]
      -P_{\xi\, ab}\left[\mathbf{E}^{a}\wedge
      e^{b}+\mathcal{D}\mathbf{E}^{ab}\right]
      \right.
    \\
    & \hspace{.5cm}
      \left.
      +d\left[\mathbf{\Theta}(e, \omega,\delta_{\xi} \omega)
      +\mathbf{E}_{a}\xi^{a} +\mathbf{E}_{ab}P_{\xi}{}^{ab} \right]\right\}.
  \end{aligned}
\end{equation}
Then, we obtain two Noether identities:\footnote{Noether identities are
  guaranteed to be satisfied off-shell, but their proofs provide good tests
  of our calculations and our understanding. Using the identities
  \begin{subequations}
    \begin{align}
      \mathcal{D}e^{a}+ T^{a}
      & =
        0,
      \\
      R^{a}{}_{b}\wedge e^{b}-\mathcal{D}T^{a}
      & =
        0,
      \\
      \mathcal{D}R^{ab}
      & =
        0,
    \end{align}
  \end{subequations}
  the first of these identities can be cast in the form
  \begin{equation}
    -\imath_{a}\left(2X_{cdef}T^{e}\wedge e^{d}\wedge R^{ef}\right)
    =
    0,
  \end{equation}
  which is manifestly true. The second Noether identity is proportional to
  \begin{equation}
    2X_{[acde}g_{b]f} e^{c}\wedge R^{de}\wedge e^{f}
    =
    0,
  \end{equation}
  which is also manifestly true.
}
\begin{subequations}
  \begin{align}
    \mathcal{D}\mathbf{E}_{a}
    -\mathbf{E}_{c}\wedge \imath_{a}T^{c}
    +\mathbf{E}_{bc}\wedge \imath_{a}R^{bc}
    & =
      0,
    \\
    \mathbf{E}^{[a}\wedge e^{b]}
    +\mathcal{D}\mathbf{E}^{ab}
    & =
      0,
  \end{align}
\end{subequations}
and an off-shell-closed 3-form current
\begin{equation}
  \begin{aligned}
  \mathbf{J}[\xi]
  & =
  \mathbf{\Theta}(e, \omega,\delta_{\xi} \omega)
  +\mathbf{E}_{a}\xi^{a} +\mathbf{E}_{ab}P_{\xi}{}^{ab}
    +\imath_{\xi}\mathbf{L}.
  \end{aligned}
\end{equation}

The off-shell closure of $\mathbf{J}[\xi]$ implies that it is locally exact
\begin{equation}
    \mathbf{J}[\xi]
     =
      d\mathbf{Q}[\xi],
\end{equation}
where 
\begin{equation}
  \mathbf{Q}[\xi]
  =
  -X_{abcd}e^{a}\wedge e^{b}P_{\xi}{}^{cd},
\end{equation}
is the \textit{Noether--Wald charge} of this theory.

By construction
\begin{equation}
  d\mathbf{Q}[\xi]
  =
  \mathbf{J}[\xi]
  =
  \mathbf{\Theta}(e, \omega,\delta_{\xi} \omega)
  +\mathbf{E}_{a}\xi^{a} +\mathbf{E}_{ab}P_{\xi}{}^{ab}
    +\imath_{\xi}\mathbf{L},
\end{equation}
and, on-shell and for a Killing vector $k$ it is not difficult to see that it
is on-shell closed\footnote{The Lagrangian is proportional trace of the
  Einstein equations
  \begin{equation}
    e^{a}\wedge \mathbf{E}_{a}
    =
     2 X_{abcd}e^{a}\wedge e^{b}\wedge R^{cd}
     =
      2 \mathbf{L}, 
    \end{equation}
    and, therefore, the Lagrangian and its inner product with $k$ vanish
    on-shell.  }
\begin{equation}
  d\mathbf{Q}[k]
  \doteq
  0,
\end{equation}
where $\doteq$ indicates an identity which may only be satisfied on-shell.

Thus, in this case, we can identify
\begin{equation}
  \mathbf{K}[k]
  \doteq
  -\mathbf{Q}[k]
\end{equation}
where $\mathbf{K}[k]$ is the on-shell-closed \textit{generalized Komar
  charge}.\footnote{The fact that the generalized Komar charge is on-shell
  closed is often referred to as \textit{conservation}, but it is more
  appropriate to say that it satisfies (the analog of) a Gauss law. See the
  discussion in the introduction of Ref.~\cite{Ballesteros:2024prz}.}

This generalized Komar charge contains two pieces: the standard GR's Komar
charge plus a term proportional to the Barbero parameter. The term
proportional to the Barbero parameter is a sort of magnetic dual of the
standard Komar charge.

In order to show this, it is convenient to rewrite the terms appearing in
it. First, using the fact that the torsion vanishes on-shell, we rewrite
$P_{k\, ab}$ in terms of the components of the exterior derivative of the
1-form dual to the Killing vector $\hat{k}=k_{\mu}dx^{\mu}$ as
\begin{equation}
  P_{k\, ab}
  \doteq
  \mathcal{D}_{[a}k_{b]}
  =
  \tfrac{1}{2}(d\hat{k})_{ab},
\end{equation}
Then, the standard GR's Komar charge coming from the EH action is 
\begin{equation}
-\tfrac{1}{2}\varepsilon_{abcd}e^{a}\wedge e^{b}P_{k}{}^{cd}  
=
-\star d\hat{k},
\end{equation}
while the term coming from the H action is
\begin{equation}
  \alpha \eta_{ab\, cd}e^{a}\wedge e^{b}P_{k}{}^{cd}
  =
  \alpha d\hat{k}.
\end{equation}

Recovering the overall normalization, the total generalized Komar charge can
be written in the form given by Prabhu in
Ref.~\cite{Prabhu:2015vua}\footnote{the modification of teh Komar charge due
  to the Holst term has also been studied in
  Refs.~\cite{DePaoli:2018erh,Oliveri:2019gvm,Oliveri:2020xls}. See also
  Refs.~\cite{Freidel:2020xyx,Wald:1990mme,Godazgar:2020kqd}.}
\begin{equation}
  \label{eq:generalizedKomarcharge}
  \mathbf{K}[k]
  =
\frac{1}{16\pi G_{N}^{(4)}}\left\{ -\star d\hat{k} +\alpha d\hat{k}\right\}.
\end{equation}

The first term in this expression is the standard on-shell
(\textit{i.e.}~dynamically) closed Komar charge 2-form of GR
\cite{Komar:1958wp}. The second, new term is topologically closed and, if
integrated over 2-spheres it gives zero unless the expression $d\hat{k}$ is
just the local form of a closed but non-exact 2-form. The two terms are
essentially identical to those whose integrals give the electric and magnetic
charges in electromagnetism $\star F$ and $F$ with $F=dA$ locally.

\vspace{.3cm}
\noindent
\textbf{4. Generalized Komar integrals and the Smarr formula of the Taub--NUT
  solution.} The integral over the 2-sphere at infinity S$^{2}_{\infty}$ of
standard Komar charge associated to the Killing vector that generates time
translations (in a stationary solution) gives $M/2$, where $M$ is the ADM
mass. Since the second term in the generalized Komar charge
Eq.~(\ref{eq:generalizedKomarcharge}) is the magnetic dual of the standard
Komar charge, it is natural to expect that its integral will, instead, give
the magnetic dual of the ADM mass, which has been customarily identified with
the NUT charge. Thus, it is natural to test this Komar form on the Lorentzian
Taub--NUT solution \cite{Taub:1950ez,Newman:1963yy}, which can be cast in the
form\footnote{Here and in what follows we are setting $G_{N}^{(4)}=1$. } 
\begin{subequations}
\label{eq:TNsolution}
\begin{align}
ds^{2}
& =
\lambda(r) (dt +A )^{2} 
         -\lambda^{-1}(r)dr^{2} -\left(r^{2}+n^{2} \right)d\Omega_{(2)}^{2},
  \\
\lambda(r)
  & =
\frac{(r-r_{+})(r-r_{-})}{r^{2}+n^{2}},
  \\
  \label{eq:1formA}
  A
  & =
    2n\left(\cos{\theta}+s\right)\,d\varphi,
  \\
  r_{\pm}
  & =
    m\pm r_{0},
    \hspace{1cm} 
    r_{0}^{2} = m^{2}+n^{2}.
\end{align}
\end{subequations}

For $s=+1,-1$, the solution has a Misner-string singularity
\cite{Misner:1963fr} over the $z>0$ and $z<0$ axes, respectively, while for
$s=0$ there are Misner-string singularities in both axes. The value of the
parameter $s$ can be changed by a change of coordinates. However, solutions
with a different value of $s$ should not be considered as physically equivalent
because the coordinate transformations that relate them are large gauge
transformations.

Misner showed in Ref.~\cite{Misner:1963fr} how to construct a singularity-free
solution gluing the singularity-free $z\geq 0$ region of the $s=-1$ solution
to the the singularity-free $z\leq 0$ region of the $s=+1$ solution along the
hypersurface $z=0$, where the $t^{(\pm)}$ time coordinates of the $s=\mp 1$
solutions are related by the coordinate transformation
\begin{equation}
t^{(+)} = t^{(-)}+4n\varphi,  
\end{equation}
which, demands, by consistency, that the time coordinate is periodic with
period $8\pi n$. Thus, the price of removing the string singularities in this
way is the introduction of closed timelike curves among other pathologies.

It has recently been shown in Ref.~\cite{Clement:2015cxa} that the
Misner-string singularities of the original Taub--NUT solutions are actually
very mild (they are transparent to free-falling particles, for instance). This
makes it possible to study their physics consistently. For instance, it has
been shown that taking the contributions of the strings into account, one can
arrive at a consistent thermodynamic description by Euclidean
\cite{Hennigar:2019ive} or Lorentzian \cite{Bordo:2019tyh} methods. Here we
want to revisit the derivation of the Smarr formula of \cite{Bordo:2019tyh}
using the generalized Komar charge Eq.~(\ref{eq:generalizedKomarcharge}).

The Smarr formulae satisfied by the thermodynamic functions and charges of the
stationary black holes of a given theory can be derived by integrating the
(on-shell vanishing) exterior derivative of the generalized Komar charge of
the theory associated to the Killing vector $k$ that becomes null over the
horizon $k^{2}\stackrel{\mathcal{H}}{=}0$, $\mathbf{K}[k]$ over a spacelike
hypersurface $\Sigma^{3}$ connecting a section of the event horizon
(preferably the bifurcation surface $\mathcal{BH}$ in bifurcate horizons) to
the sphere at spatial infinity S$^{2}_{\infty}$ and using Stokes theorem
\cite{Bardeen:1973gs,Carter:1973rla,Magnon:1985sc,Bazanski:1990qd,Kastor:2010gq}. Since,
by construction, $\partial \Sigma^{3} = \mathcal{BH} \cup$S$^{2}_{\infty}$,
taking into account the different orientations of the two components of the
boundary, we get a relation between two generalized Komar
integrals:\footnote{The generalized Komar integral at infinity can actually be
  performed over any sphere of finite radius. Since it satisfies a Gauss law,
  it gives the same value, which must be independent of that radius.}
\begin{equation}
  0 = \int_{\Sigma^{3}}d\mathbf{K}[k]
  =
  \int_{\mathcal{BH}}\mathbf{K}[k] - \int_{\mathrm{S}^{2}_{\infty}}\mathbf{K}[k].
\end{equation}

In absence of H term ($\alpha=0$), the integral over the bifurcation surface
is naturally expressed in terms of thermodynamical properties of the horizon:
temperature $T$ and entropy $S$ in the case at hands (pure GR). The result is
just $ST$.  The integral at infinity is naturally expressed in terms of the
conserved charges associated to $k$: ADM mass $M$ and angular momentum
$J$. The result is $M/2 -\Omega J$ where $\Omega$ is the angular velocity of
the horizon.

Substituting in the above identity we get the standard $\alpha=0$ Smarr
formula
\begin{equation}
  \label{eq:Smarr0}
M/2 -\Omega J -ST =0.  
\end{equation}

As noticed in Ref.~\cite{Bordo:2019tyh}, in the $s=\pm 1,0$ Taub--NUT
spacetimes ($\Omega=J=0$), the boundary of $\Sigma^{3}$ contains additional
contributions associated to the string singularities and the Smarr
formula\footnote{It has been recently shown in Ref.~\cite{Barbagallo:2025tjr}
  that, if we use instead Misner's construction, the string singularities
  unavoidably reappear in the spacelike hypersurfaces. Thus, we must always
  consider their contribution.}
\begin{equation}
  \int_{\mathcal{BH}}\mathbf{K}[k]
  -\int_{\mathrm{S}^{2}_{\infty}}\mathbf{K}[k]
  +\int_{\rm string+}\mathbf{K}[k]
  -\int_{\rm string-}\mathbf{K}[k]
  =
  0,
\end{equation}
where string $\pm$ stands for the strings in the $\pm z>0$ semiaxes, when they
are present.

The integral over the string in the $\pm z>0$ semiaxes are naturally expressed
in terms of the \textit{Misner potentials} of the strings $\psi_{\pm}$ times a
conjugate charge $\mathcal{N}_{\pm}$ (the \textit{Misner string strength}) and
one arrives at the $\alpha=0$ Smarr formula \cite{Bordo:2019tyh}
\begin{equation}
  \label{eq:Smarrstrings}
M/2 -ST -\psi_{+}\mathcal{N}_{+} -\psi_{-}\mathcal{N}_{-}=0,  
\end{equation}
that can be checked by using the explicit values of the variables
\begin{subequations}
  \begin{align}
    M
    & =
      m,
      \\
    T
    & =
      \frac{\kappa}{2\pi}
      =
      \frac{1}{4\pi r_{+}},
    \\
    S
    & =
      \pi(r^{2}_{+}+n^{2}),
    \\
    \psi_{\pm}
    & =
      \frac{\kappa_{\pm}}{4\pi}
      =
      \frac{1}{8\pi n(1\pm s)},
    \\
    \mathcal{N}_{\pm}
    & =
      -\frac{2\pi n^{3}(1\pm s)^{2}}{r_{+}}.
  \end{align}
\end{subequations}

The presence of the H term ($\alpha\neq 0$) does not modify the Smarr formula
Eq.~(\ref{eq:Smarrstrings}) because the standard homogeneity arguments tell us
that there will not be a new term proportional to $\alpha$.\footnote{There
  will be a term of the form $\Phi_{\alpha}\delta \alpha$ in the right-hand
  side of the first law, though. The mechanism is the same that, in theories
  containing scalar fields, leads to terms of the form
  $\Sigma\delta \phi_{\infty}$ \cite{Gibbons:1996af}.} The values of charges
and thermodynamic potentials as functions of the integration constants $m$ and
$n$ may be modified, though. For instance, the ADM mass will still be given by
the integral
\begin{equation}
  M
  \equiv
  \frac{1}{8\pi}\int_{\mathrm{S}^{2}_{\infty}} \mathbf{K}[k].
\end{equation}
However, $\mathbf{K}[k]$ contains a new term and the relation between the ADM
mass and the parameter $m$ in the solution Eq.~(\ref{eq:TNsolution}) will
differ by a term proportional to $\alpha$. The same will happen to other
physical quantities of the Taub--NUT solution.

Since the term proportional to $\alpha$ in $\mathbf{K}[k]$ is closed by
itself, we can simply compute the integrals of that term over the same
components of the boundary. The sum of these contributions vanishes by itself
(Stokes theorem) and the Smarr formula Eq.~(\ref{eq:Smarrstrings}) will be
satisfied once again.

Let us first consider the integral at spatial infinity. We can use this
integral as the definition of the NUT charge $N$:
\begin{equation}
  \label{eq:NUTchargedef} 
  N
  \equiv
  -\frac{1}{8\pi}\int_{\mathrm{S}^{2}_{\infty}}d\hat{k}.  
\end{equation}
For $k=\partial_{t}$, the integrand is the pullback of
$d\hat{k}=\lambda' dr\wedge (dt+A) +\lambda dA$ and, then, at spatial infinity
$d\hat{k}\to dA$. The integral of $dA$ over the 2-sphere (at infinity or
anywhere else) is identical to the integral of the electromagnetic field of a
Dirac monopole over a 2-sphere and it can be performed in the same fashion: if
we use the Wu--Yang description, we integrate
$2nd\left[(\cos{\theta}-1) d\varphi\right]$ over the string-free
$\theta\leq \pi/2$ hemisphere \textit{and} that of
$2nd\left[(\cos{\theta}+1) d\varphi\right]$ over the string-free
$\theta \geq \pi/2$ hemisphere and sum the results. If we use the Dirac
description, we integrate $2nd\left[(\cos{\theta}-1)d\varphi\right]$
over the whole 2-sphere minus the region $\theta = \pi-\epsilon$ surrounding
the string lying in the $z<0$ semiaxis \textit{or} we integrate
$2nd\left[(\cos{\theta}+1)d\varphi\right]$ over the whole 2-sphere
minus the region $\theta = \epsilon$ surrounding the string lying in the $z>0$
semiaxis \textit{or} we integrate
$2nd\left(\cos{\theta}\wedge d\varphi\right)$ over the whole 2-sphere minus
the regions $\theta = \epsilon$ and $\theta = \pi-\epsilon$ both strings and
we let $\epsilon\to 0$ in the result. Either way, we get
\begin{equation}
  N
  =
  n,
\end{equation}
so the ADM mass now becomes
\begin{equation}
  M
  =
  m-\alpha n = m-\alpha N.
\end{equation}

This result gives rise to a gravitational version of the Witten effect
\cite{Witten:1979ey}: the Taub-NUT solution with $m=0$ has a non-vanishing ADM
mass $M=-\alpha N$.\footnote{A similar effect may be produced by the
  introduction of a topological Euler-density term in the action
  \cite{Foda:1984gq}. The Euler-density term would be the analog of the
  $\theta$-term if the EH term was the analog of the standard Yang--Mills
  kinetic terms, which is not, but it leads to similar results. The combined
  effect of the H and Euler-density terms has been considered in
  \cite{Durka:2011zf,Durka:2019ajz}.}

Let us now compute the other integrals.

The integral over the bifurcation sphere of $d\hat{k}$ vanishes identically
because $\lambda\stackrel{\mathcal{H}}{=}0$ and $dr \stackrel{\mathcal{H}}{=}0$:
\begin{equation}
  \frac{1}{8\pi}\int_{\mathcal{BH}}d\hat{k}
  =
  0.
\end{equation}

When there is a string lying on the $z>0$ semiaxis ($s=+1$), the contribution
of the string must be computed as the integral over the cone $t=$ constant,
$\theta = \epsilon$ of the pullback of
$d\hat{k}=\lambda' dr\wedge (dt+A) +\lambda dA$, with
$A=2n(\cos{\theta}+1)d\varphi$ from $r=r_{+}$ to $r=\infty$:
\begin{equation}
  \begin{aligned}
  \frac{1}{16\pi}\int_{\rm string+}d\hat{k}
  & =
  \frac{1}{16\pi}\int_{\rm string+}
    \left[\lambda' dr\wedge A +\lambda dA \right]
    \\
  & =
  \frac{1}{16\pi}\int_{\rm string+}
    \left[\lambda' 2n(\cos{\theta}+1) dr\wedge d\varphi
    -2n \lambda \sin{\theta}d\theta\wedge d\varphi \right]
    \\
  & =
  \frac{n}{8\pi}(\cos{\epsilon}+1) \int_{\rm string+}d\lambda\wedge d\varphi
    \\
  & =
    \frac{n}{8\pi}(\cos{\epsilon}+1)
    \int_{\lambda=0}^{\lambda=1}d\lambda \int_{\varphi=0}^{\varphi=2\pi}d\varphi
    \\
  & =
  \frac{n}{4}(\cos{\epsilon}+1) 
    \\
  & \longrightarrow 
  \frac{n}{2}.
  \end{aligned}
\end{equation}

The same result is obtained when the string lies in the $z<0$ semiaxis or when
there are two strings, if one takes into account correctly the
orientation. Thus, the values of the terms $\psi_{\pm}\mathcal{N}_{\pm}$ are
shifted by $\alpha n/2$ and, since, due to their definition,  $\psi_{\pm}$
should not be affected by the new term in the action, the change must be due
to the change in the value of $\mathcal{N}_{\pm}$:
\begin{equation}
  \label{eq:Npmalpha}
      \mathcal{N}_{\pm}
    =
      -\frac{2\pi n^{3}(1\pm s)^{2}}{r_{+}}+\alpha 4\pi n^{2}(1\pm s).
\end{equation}

As we have briefly mentioned before, now the entropy is a homogeneous function
$S(M,\mathcal{N}_{\pm},\alpha)$ and, although the Smarr formula does not
acquire any terms proportional to $\alpha$, there will be a new term in the
first law
\begin{equation}
  \delta M
  =
  T\delta S +\psi_{+}\delta \mathcal{N}_{+}+\psi_{-}\delta \mathcal{N}_{-}
  +\Phi_{\alpha}\delta \alpha\,
\end{equation}
where the new chemical potential $\Phi_{\alpha}$ is defined by 
\begin{equation}
  \Phi_{\alpha}
  =
  T\frac{\partial S}{\partial \alpha}.
\end{equation}

The calculation of that partial derivative is quite involved, since we must
take into account that both $m$ and $n$ are functions of the thermodynamical
variables $M,\mathcal{N}_{\pm}$ and, in particular, of  $\alpha$. Thus,
\begin{equation}
  \frac{\partial m}{\partial \alpha}
  =
  \frac{\partial (M-\alpha n)}{\partial \alpha}
  =
  -n -\alpha \frac{\partial n}{\partial \alpha},
\end{equation}
and $\partial n/\partial \alpha$ can be found by taking a partial derivative
with respect to $\alpha$ in both sides of Eq.~(\ref{eq:Npmalpha}), using the
above result and solving for $\partial n/\partial \alpha$. The final expression
for $\Phi_{\alpha}$ is a complicated and not too illuminating function of $m,n$
and $\alpha$.

\vspace{.3cm}
\noindent
\textbf{5. Discussion.} In this paper we have shown that the strong
parallelism existing between the Holst term in first-order gravity and the
$\theta$-term in Yang--Mills theories leads to a gravitational analog of the
Witten term sourced bow by the NUT charge, for which there is a very natural
definition Eq.~(\ref{eq:NUTchargedef}) that strengthens its interpretation as
a ``magnetic mass.'' 

It would be interesting to apply these ideas to find a good definition of the
topological charge carried by the The Gross-Perry--Sorkin KK monopole (or
Euclidean Taub--NUT solution) \cite{Gross:1983hb,Sorkin:1983ns}. Work in this
direction is currently underway.

\vspace{.3cm}
\noindent
\textbf{Acknowledgments.} The work of JLV-F and TO has been supported in part
by the MCI, AEI, FEDER (UE) grants PID2021-125700NB-C21 (``Gravity,
Supergravity and Superstrings'' (GRASS)) and IFT Centro de Excelencia Severo
Ochoa CEX2020-001007-S.  The work of JLVC has been supported by the CSIC
JAE-INTRO grant JAEINT-24-02806. TO wishes to thank M.M.~Fern\'andez for her
permanent support.

\vspace{.3cm}
\noindent
\textbf{Appendix} Under an infinitesimal GCT generated by the vector field
$\xi$ the metric transforms as
\begin{equation}
  \delta_{\xi}g_{\mu\nu}
  =
  -\mathcal{L}_{\xi}g_{\mu\nu}
  =
  -\xi^{\rho}\partial_{\rho}g_{\mu\nu}-2\partial_{(\mu}\xi^{\rho}g_{\nu)\rho}.
\end{equation}

The partial derivatives can be replaced by the Levi-Civita covariant
derivative, which gives the standard expression
\begin{equation}
  \delta_{\xi}g_{\mu\nu}
  =
-2\nabla^{L-C}_{(\mu}\xi_{\nu)},
\end{equation}
but, when we use a metric-compatible but torsionful connection
we arrive to the expression
\begin{equation}
  \delta_{\xi}g_{\mu\nu}
  =
-2\nabla_{(\mu}\xi_{\nu)} +2\xi^{\rho}T_{\rho(\mu\nu)},
\end{equation}
and the Killing vector equation takes the form
\begin{equation}
  \label{eq:torsionfulKVE}
  \nabla_{(\mu}k_{\nu)} -k^{\rho}T_{\rho(\mu\nu)}
  =
  0.
\end{equation}

In this context the momentum map associated to the vector $\xi$ tat enters in
the definition of the parameters of the compensating Lorentz transformations
$\sigma_{\xi}{}^{ab}$ takes the form
\begin{equation}
  \label{eq:momentummap}
  P_{\xi\, ab}
  =
  \mathcal{D}_{[a}\xi_{b]} -\xi^{c}T_{c[ab]},
\end{equation}
and
\begin{equation}
  \label{eq:sigmaxiab}
  \sigma_{\xi}{}^{ab}
  =
  \imath_{\xi}\omega^{ab} -P_{\xi}{}^{ab},
\end{equation}
where $\omega^{ab}$ is the (torsionful) lorentz (spin) connection.

Following the general recipe, the transformation of the Vielbein under GCTs is
given by
\begin{equation}
  \label{eq:deltaxiea}
  \delta_{\xi}e^{a}
  =
  -\mathcal{L}_{\xi}e^{a} + \sigma_{\xi}{}^{a}{}_{b}e^{b}
  =
  -\left(\mathcal{D}\xi^{a} -\imath_{\xi}T^{a}
    +P_{\xi}{}^{a}{}_{b}e^{b}\right),
\end{equation}
and it is not difficult to see that it vanishes identically for Killing
vectors because it is proportional to the Killing vector equation
(\ref{eq:torsionfulKVE}).

Using the same rule we find 
\begin{equation}
  \label{eq:deltaxioab}
      \delta_{\xi}\omega^{ab}
    =
    -\left(\imath_{\xi}R^{ab}+\mathcal{D}P_{\xi}{}^{ab}\right).
\end{equation}
In the torsionless case, one can show that this expression vanishes
identically for Killing vectors because it is the integrability condition of
the Killing vector equation or, equivalently, of the condition
$\delta_{k}e^{a}=0$. In the torsionful case, the integrability condition of
$\delta_{k}e^{a}=0$ with $\delta_{\xi}e^{a}$ given by Eq.~(\ref{eq:deltaxiea})
is
\begin{equation}
  \mathcal{D}\delta_{k}e^{a}
  =
  -\left(\imath_{k}R^{a}{}_{b}+\mathcal{D}P_{k}{}^{a}{}_{b}\right)\wedge e^{b}
  -\delta_{k}T^{a},
\end{equation}
where we have defined, following the general rule,\footnote{Observe that this
  rule is consistent with
  \begin{subequations}
    \begin{align}
      T_{\mu\nu}{}^{a}
      & =
        T_{\mu\nu}{}^{\rho}e^{a}{}_{\rho},
      \\
      \delta_{\xi}T_{\mu\nu}{}^{\rho}
      & =
        -\mathcal{L}_{\xi}T_{\mu\nu}{}^{\rho},
      \\
      \delta_{\xi}e^{a}
      & =
        -\mathcal{L}_{\xi}e^{a} + \sigma_{\xi}{}^{a}{}_{b}e^{b}.
    \end{align}
  \end{subequations}
}
\begin{equation}
  \delta_{\xi}T^{a}
  =
  -\mathcal{L}_{\xi}T^{a} +\sigma_{\xi}{}^{a}{}_{b}T^{b},
\end{equation}
where $\sigma_{\xi}{}^{ab}$ is the parameter of the compensating or induced
local Lorentz transformation in Eq.~(\ref{eq:sigmaxiab}).

Thus, $\delta_{k}e^{a}=0$ implies $\delta_{k}\omega^{ab}=0$ if
$\delta_{k}T^{a}=0$. Since the torsion is an independent field, it is clearly
necessary to demand its invariance as one of the conditions that ensure that
$\delta_{k}$ is a symmetry of all the fields of the theory. In other words: we
must demand Killing vectors which also leave the torsion field invariant.

Finally, we notice that the same rule leads to a transformation of the
curvature consistent with the Palatini identity
\begin{equation}
       \delta_{\xi}R^{ab}
       =
       \mathcal{D}\delta_{\xi}\omega^{ab}
       =
    -\mathcal{D}\left(\imath_{\xi}R^{ab}+\mathcal{D}P_{\xi}{}^{ab}\right),
\end{equation}
which vanishes identically for Killing vectors that leave invariant the
torsion field.


\end{document}